# MECHANICAL/ELECTRICAL PROPERTIES OF MWCNT/PP FILMS FOR STRUCTURAL HEALTH MONITORING OF GF/PP JOINTS

W. Li[a] and G. Palardy[a*]

[a] *Department of Mechanical and Industrial Engineering, Louisiana State University*
*3261 Patrick F. Taylor Hall*
*Baton Rouge, LA 70803*

## Abstract

While welding of thermoplastic composites (TPCs) is a promising rivetless method to reduce weight, higher confidence in joints' structural integrity and failure prediction must be achieved for widespread use in industry. In this work, we present an innovative study on damage detection for ultrasonically welded TPC joints with multi-walled carbon nanotubes (MWCNTs) and embedded buckypaper films. MWCNTs show promise for structural health monitoring (SHM) of composite joints, assembled by adhesive bonding or fusion bonding, through electrical resistance changes. This study focuses on investigating multifunctional films and their suitability for ultrasonic welding (USW) of TPCs, using two approaches: 1) MWCNT-filled polypropylene (PP) nanocomposites prepared via solvent dispersion, and 2) high conductivity MWCNT buckypaper embedded between PP films by hot pressing. Nanocomposite formulations containing 5 wt% and 10 wt% MWCNTs were synthesized using solvent dispersion method, followed by compression molding to manufacture films. The effect of MWCNT concentration on electrical and dynamic mechanical behavior of multifunctional films was examined with a Sourcemeter and Dynamic Mechanical Analyzer, and a comparison was made between 5 - 20 wt% MWCNT/PP films based on previous research. Glass fiber/polypropylene (GF/PP) composite joints were ultrasonically welded in a single lap shear configuration using buckypaper and MWCNT/PP films. Furthermore, electrical resistance measurements were carried out for joints under bending loads. It was observed that 15 wt% and 20 wt% MWCNT/PP films had higher stability and sensitivity for resistance response than embedded buckypaper and films with low MWCNT contents, demonstrating their suitability for USW and potential for SHM.

## 1. Introduction

During the last decades, thermoplastic materials have been used in several industries due to their outstanding advantages, such as low weight, fast processing, recyclability, and corrosion resistance compared to conventional metals and alloys [1, 2]. In particular, the applications of polypropylene (PP), belonging to polyolefins and one of the most extensively produced semi-crystalline thermoplastic polymers, dramatically increased in the past years, ranging from automotive, aircraft, military, construction, and medical care to appliances, furniture, and clothing [2, 3]. These applications are based on PP's processability, excellent physical and chemical properties, and low cost [4, 5]. However, there are still some limitations that hinder the application of PP, including poor electrical, mechanical, and thermal properties [2, 6].

To overcome these limitations, nano fillers, such as carbon nanotubes (CNTs), carbon nanofibers (CNFs), carbon black, and graphene are incorporated into polymer matrices [7-10]. In particular, multi-walled carbon nanotubes (MWCNTs) have shown potential as ideal nanofillers for improving properties of polymers in recent years. For instance, Wang et al. [11] pre-dispersed MWCNTs into PP by high-speed rotating before melt extrusion, and they found that the mechanical and electrical properties of the resulting nanocomposites were significantly improved. Zetina-Hernandez et al. [10] studied 4 - 10 wt% MWCNT/PP nanocomposites and concluded that the electrical conductivity and elastic modulus both increased with MWCNT weight fraction. Salih Hakan [2] investigated the mechanical, thermal, and rheological properties of PP composites filled with MWCNTs. Mechanical test results showed that the tensile strength and flexural strength increased due to the addition of MWCNTs. Moreover, the thermal and rheological behavior of PP was enhanced as well.

Incorporation of CNTs into a polymer matrix creates an electrical network, allowing strain sensing and damage detection by electrical resistance changes, a promising structural health monitoring (SHM) technique that has been studied for fiber-reinforced polymers and adhesive joints. For example, Augustin et al. [12] used single-walled carbon nanotubes (SWCNTs)/epoxy-based adhesive films to monitor the impact damage in adhesively bonded glass fiber-reinforced polymer (GFRP) joints through in-plane and through-thickness electrical resistance measurements, showing potential for SHM. They also detected crack initiation and growth of adhesively bonded carbon fiber-reinforced polymer (CFRP) joints during cyclic loading with SWCNT-modified epoxy-based adhesive films through electrical resistance changes [13]. Sanchez-Romate et al. [14] incorporated CNTs into adhesive films to monitor the crack propagation of adhesively bonded composite joints subjected to peeling by means of electrical resistance measurements, showing a high potential for SHM. In addition to composite joints, Mactabi et al. [15] prepared aluminum joints using CNT reinforced adhesive and explored the damage of adhesive joints under fatigue loading using electrical resistance measurements. They reported that CNT networks exhibited a high sensitivity to joint fracture. Apart from CNTs, buckypaper, a thin, paper-like membrane with well-controlled dispersion of CNTs, CNFs, or other carbon-based nanomaterials [16], demonstrated promise as multifunctional sensors, including their use in composite laminates [16-18].

Unlike SHM of adhesively bonded joints, which has been investigated by several researchers, damage detection of welded thermoplastic composite (TPC) joints has not been studied. One particular challenge includes development of sensing films compatible with the welding process. Ultrasonic welding (USW), a joining method discussed in this research, was shown to create strong TPC joints. However, to further support its widespread application in industry, better understanding and monitoring of joints' structural integrity during use should be achieved. In our previous work, we successively evaluated the capability of MWCNT/PP films for ultrasonic welding and damage detection of glass fiber/polypropylene (GF/PP) welds under tension loading [19]. In this study, our objective is to further explore multifunctional films and their potential for SHM for TPC joints subjected to bending loads.

Based on the aim of this work, various MWCNT/PP films were compared. Films were made from commercial masterbatches, solvent-based dissolution method, and buckypaper embedded between pure PP films. The electrical properties of 5 wt%, 10 wt%, 15 wt%, and 20 wt% MWCNT/PP films were estimated by measuring their electrical conductivity. Next, the effect of MWCNT weight fraction on dynamic mechanical properties (storage and loss moduli) were explored to assess heat generation behavior for ultrasonic welding. The welding process for GF/PP joints with MWCNT/PP films and buckypaper films was investigated. Finally, the potential of various films for structural health monitoring of welded TPC joints was assessed under bending loads.

## 2. Materials and Methods

### 2.1 Materials

To manufacture films with high MWCNT weight fractions (> 10 wt%), PP masterbatches containing 15 wt% and 20 wt% MWCNTs were bought from Cheap Tubes Inc. (Grafton, VT, USA). For nanocomposite films containing lower MWCNT weight fractions, MWCNT powders with a purity above 95 wt% were used (Cheap Tubes Inc., Grafton, VT, USA). Their average outer diameter and length are 10-20 nm and 10-30 µm, respectively. They have a specific surface area of at least 233 $m^2$/g and a density approximately equal to 2.1 g/$cm^3$ at room temperature. PP pellets with a nominal granule size of 5 mm and a melt flow rate of 6 g/min, obtained from Goodfellow (Huntingdon, Cambridgeshire, England), were used as the polymer matrix. As a solvent, p-Xylene (1,4-Dimethylbenzene), anhydrous (≥99%), with a density of 0.861 g/ml and a boiling point of 138 ℃, was purchased from Sigma-Aldrich (St. Louis, MO, USA). MWCNT blend buckypaper with a surface electrical resistivity of 1.5 Ohm/square was purchased from NanoTechLabs (Yadkinville, NC, USA).

To manufacture the composite adherends, unidirectional glass fiber/polypropylene (GF/PP) prepregs, IE 6030, were supplied by PolyOne (Avon Lake, OH, USA) with a thickness of 0.33 mm.

### 2.2 Sample Preparation

#### 2.2.1 Manufacture of MWCNT/PP Films

Dispersion state of MWCNTs in polymers plays an important role in the properties of nanocomposites, especially for electrical properties [20]. To improve dispersion of MWCNTs, chemical and physical methods were employed in combination when fabricating nanocomposites containing 5 wt% and 10 wt% MWCNTs. As shown in Figure 1, 0.1 g MWCNTs (> 95 wt%) and 2 ml xylene were mixed into a beaker and placed in an ultrasonic bath sonicator (Symphony™ 97043-988, VMR International, Radnor, PA, USA) for 6 h. Then, 1.9 g PP pellets and 38 ml xylene were heated at 210 ℃ for 1.5 h with a magnetic stirring bar using a Corning® PC-420D stirring hot plate (Glendale, AZ, USA) until completely dissolved. The MWCNT/xylene mixture was poured into the PP/xylene beaker on the hot plate at 210 ℃ to continue stirring for 1 h. After achieving uniform dispersion, the MWCNTs dispersed PP suspension was dried in an oven at 80 ℃ for 24 h to remove the xylene solvent. Finally, 5 wt% MWCNT/PP nanocomposites were obtained. Similarly, 10 wt% MWCNT/PP nanocomposites were manufactured (MWCNTs: xylene = 0.2 g: 4 ml, PP: xylene = 1.8 g: 36 ml).

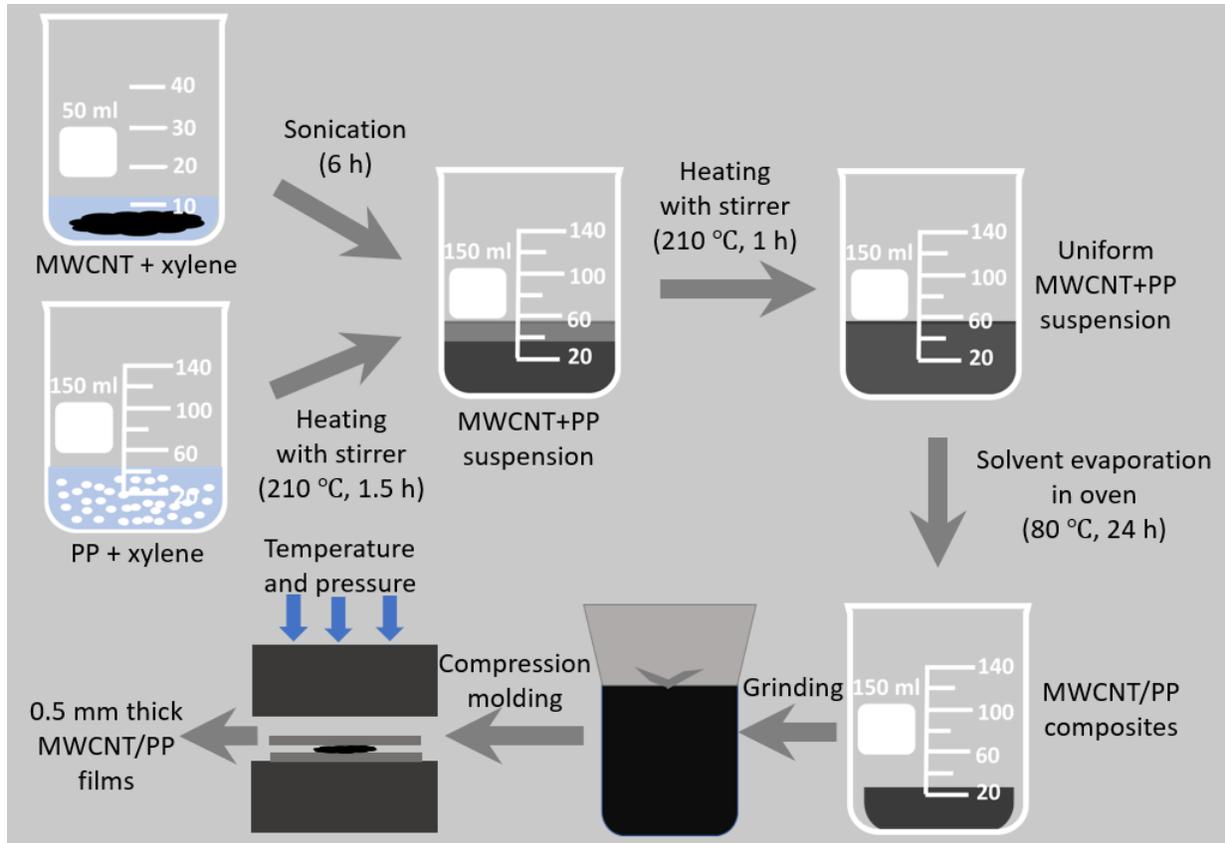

*Figure 1. Schematic of fabrication process of MWCNT/PP nanocomposites.*

The dried 5 wt% and 10 wt% MWCNT/PP nanocomposites were ground to fine powders using a grinder, followed by compression molding through a heated laboratory press (DAKE, Grand Haven, MI, USA) to fabricate 5 wt% and 10 wt% MWCNT/PP films at 180 ℃ for 15 min with a pressure of 0.8 Mpa. During molding, 0.5 mm-thick stainless-steel shims were placed around the powders to maintain the films' final thickness. To manufacture the high MWCNT content nanocomposite films (15 wt% and 20 wt%), pellets from the commercial masterbatches were similarly placed in the heated press.

For the buckypaper PP films, buckypaper with a thickness around 0.12 mm was embedded between two 0.5 mm-thick pure PP films. The stack was heat pressed at 150 ℃ for 15 min under 10 tons. Finally, a film with a total thickness approximately equal to 1.12 mm was obtained.

### 2.2.2 Manufacture of GF/PP Laminates

Eight layers (254 mm × 254 mm) of GF/PP prepregs were stacked in the 0° direction and compression-molded at 180 ℃ for 15 min under 1 MPa. The obtained laminate with a thickness of approximately 1.8 mm was cut into 101.6 mm × 25.4 mm rectangular strips using a PICO 155 Precision Saw (PACE Technologies, Tucson, AZ, USA). The long side of the rectangle was aligned with the direction of the glass fibers.

## 2.3 Characterization

### 2.3.1 Electrical Conductivity Measurements

The electrical resistance of 50 mm × 15 mm × 0.5 mm specimens, cut from the center of the 5 wt%, 10 wt%, 15 wt%, and 20 wt% MWCNT/PP films after compression molding, were measured with a Keithley Sourcemeter 2604B (Cleveland, Ohio, USA) at room temperature. Samples were sandwiched between two copper electrodes (25.4 mm × 25.4 mm × 3.2 mm), and six DC voltages of 1 V, 2 V, 4 V, 6 V, 8 V, and 10 V were applied to gather resistance using KickStart software, as illustrated in Figure 2, based on our previous study [21]. Five replications were done for each weight fraction and their average was used. The electrical conductivity of films, $\sigma$, was calculated as:

$$\sigma = \frac{L}{RA} \tag{1}$$

where $L$ is the length of films between copper electrodes, $R$ is the measured resistance, and $A$ is the cross-sectional area of the films.

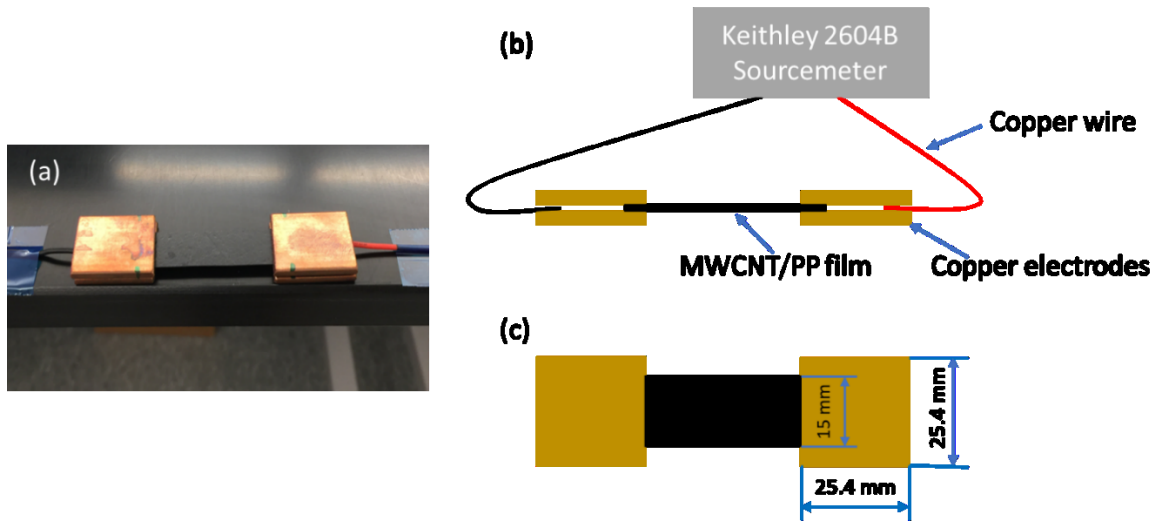

*Figure 2. (a) MWCNT/PP films and copper electrodes placement for electrical conductivity measurements; (b) side view of (a); (c) top view of (a).*

### 2.3.2 Dynamic Mechanical Analysis (DMA)

The effect of MWCNT concentration on dynamic mechanical properties of MWCNT/PP nanocomposite films was investigated using a Q800 DMA (TA instruments, New Castle, DE, USA). 18 mm × 8 mm × 0.5 mm films were cut. A QDRIVER3 adjustable torque screwdriver with 5 in-lb torque was used to tighten samples. The temperature scan was performed from -50 ℃ to 120 ℃ at a heating rate of 3 ℃/min and a frequency of 1 Hz in a continuous nitrogen supply environment.

### 2.3.3 Ultrasonic Welding

Two 101.6 mm × 25.4 mm rectangular GF/PP strips were welded in a single lap configuration with an overlap of 25.4 mm × 12.7 mm using a Dynamic 3000 ultrasonic welder (Rinco Ultrasonics, Danbury, CT, USA), under a frequency of 20 kHz, with a 40 mm diameter sonotrode. The booster gain and horn gain were 1: 1.5 and 1: 3.85, respectively. A vibration amplitude of 38.1 µm was applied, and a 60% travel value (the percentage of the film thickness) was chosen based on our previous research [19]. After the vertical displacement of sonotrode reached 60% travel, a solidification force of 1000 N was applied for 4000 ms. Figure 3 shows a detailed description of the sample installation. Top adherend, supported by a GF/PP strip, and bottom adherend were respectively fixed by two custom-made clamps with 25.4 mm-wide grooves to keep them from moving during welding. A PP film (as a reference), MWCNT/PP film, or buckypaper film with an area larger than the overlap area was sandwiched between the two adherends.

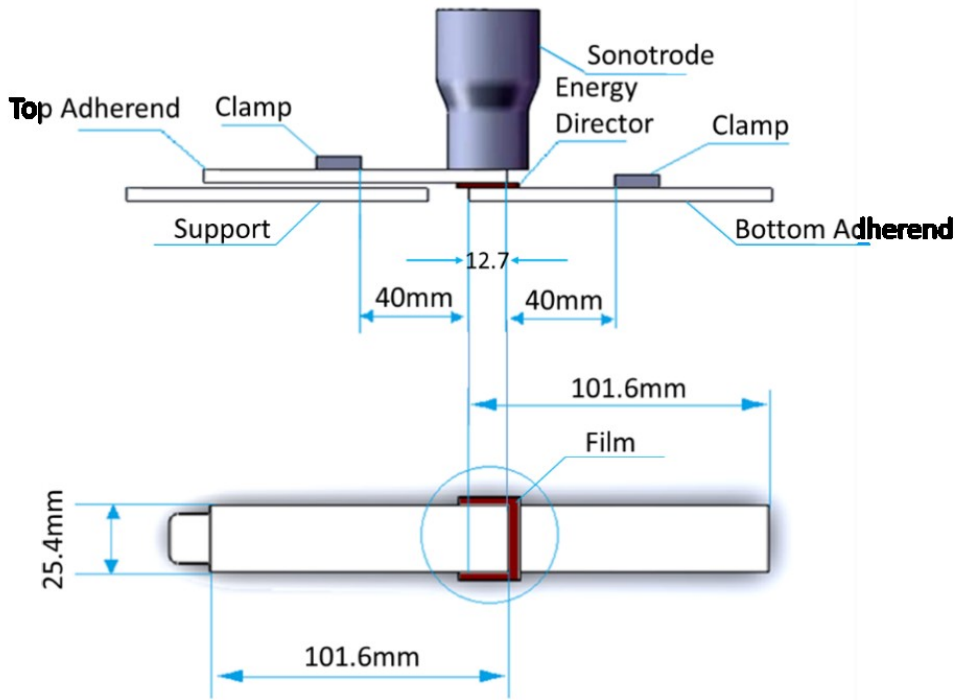

Figure 3. Schematic of welded samples (reproduced from [19] with permission).

### 2.3.4 Bending Tests

After welding, polymer flash at the edges of the joints welded with 5 wt%, 10 wt%, 15 wt%, and 20 wt% MWCNT/PP films, and excess buckypaper were removed. SPI #05002-AB silver conductive paint (Structure Probe, Inc., West Chester, PA, USA), reducing contact resistance, was brushed on both sides of the overlap along the long side of the adherends. Two 30 AWG copper wires were placed on the silver coating, and then painted again. All samples were dried overnight in a fume hood. To assess potential of the nanocomposite films for monitoring during bending loads, a simple experimental setup, as shown in Figure 4, was used. The welded sample was placed about 33 mm from the center of the overlap with a C-clamp, and a downward load (~ 14 N) was periodically applied to the other end of the beam using a M2-100 series M2 digital force gauge (Mark-10, Copiague, NY, USA). The load was applied at intervals of 35 s, 20s, 10s, or 6 s for a bending duration of 3 s, resulting in a maximum vertical displacement of 45 mm. A 1 mA current was applied, and the resistance of the joints was measured by a Keithley Sourcemeter 2604B (Cleveland, Ohio, USA). Five samples were tested for each condition and nanocomposite film.

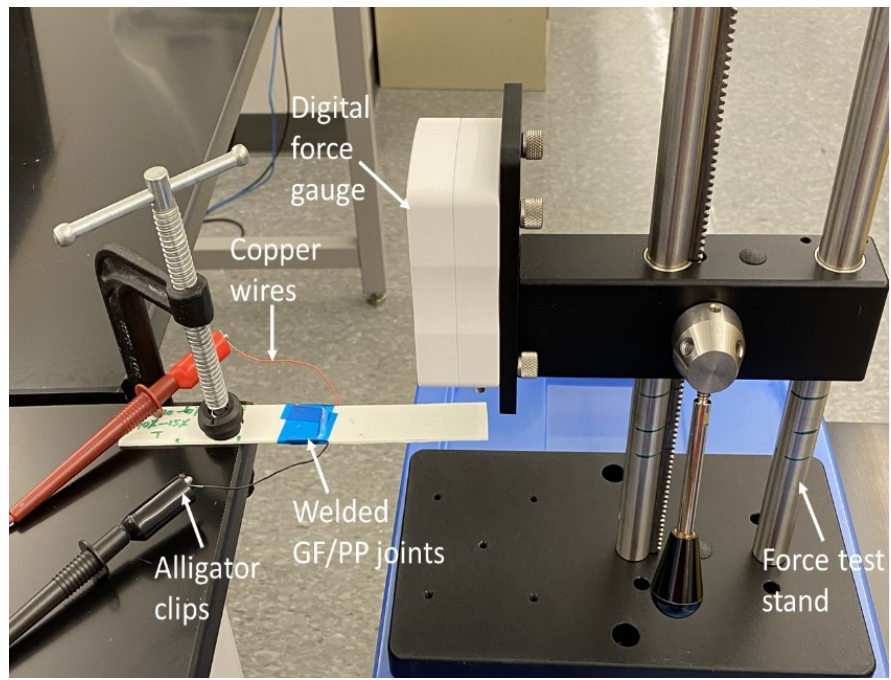

*Figure 4. Bending test setup.*

## 3.  Results and Discussion

### 3.1  Electrical Properties

Figure 5 shows the electrical conductivity of 5 wt%, 10 wt%, 15 wt%, and 20 wt% MWCNT/PP films as a function of voltage. Overall, the electrical conductivity values measured for this range of CNT weight fractions are similar to results presented in the literature [10]. The electrical conductivity starts to increase more significantly, especially for films with high MWCNT concentration, when voltage increases above 2 V. In addition, electrical conductivity increases with the amount of MWCNTs up to 15 wt%, and then decreases. It is expected that the films with a MWCNT concentration lower than 15 wt%, which have significantly lower electrical conductivity, may possess fewer paths for the conductive network between MWCNTs [8]. The lower conductivity of the films with high MWCNT concentration (20 wt%) may be associated with the poor dispersion of MWCNTs due to the formation of clusters, which hinders connections between MWCNTs [21]. Therefore, 15 wt% MWCNT/PP films show greater potential for stable structural health monitoring of the welded joints using resistance changes.

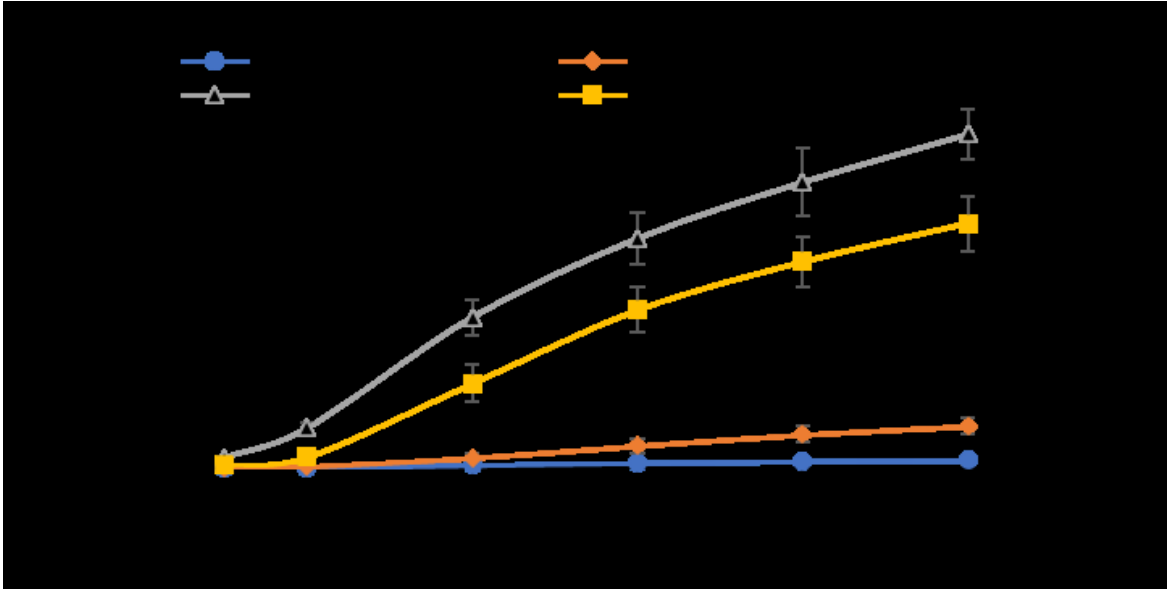

*Figure 5. The electrical conductivity of 5 wt%, 10 wt%, 15wt%, and 20 wt% MWCNT/PP films as a function of voltage.*

## 3.2 Dynamic Mechanical Analysis

Figure 6a plots the change of storage modulus for pure PP, 5 wt%, 10 wt%, 15 wt%, and 20 wt% MWCNT/PP films during temperature sweep. The storage modulus curves of pure PP and 5 wt% MWCNT/PP films almost overlap, suggesting this weight fraction did not significantly affect the elastic behavior of the films. Films containing higher MWCNTs fractions (10 to 20 wt%) have a higher storage modulus, especially at low temperature. This may be caused by the interaction between MWCNTs and PP, resulting in more connected networks and MWCNTs adsorbing on the surface of PP chains, which restrain their movement [22]. This increasing behavior shows that MWCNTs have the ability to reinforce PP films. After reaching the glass transition temperature ($T_g$), -10 ℃ ~ 10 ℃ for PP [3], storage modulus drops dramatically.

Similarly, the loss modulus generally increases with MWCNT concentration, presented in Figure 6b as a function of temperature. High MWCNTs weight fractions can impact the viscous behavior of PP, but at 20 wt%, some aggregates or clusters may contribute to lower loss modulus values. Furthermore, at low temperature, the loss modulus, energy dissipating with heat, is relatively stable, and a peak representing the maximum heat dissipation per unit deformation [22] appears around 7 ℃ in the glass transition region. As the temperature exceeds the glass transition temperature, less energy is dissipated on account of the reduced intermolecular friction, so loss modulus decreases. More importantly, during ultrasonic welding process, the loss modulus plays a role in viscoelastic heat generation. The loss modulus of nanocomposite films increases significantly with the MWCNT loading up to 20 wt% with respect to pure PP films. Therefore, that these multifunctional films should result in a faster viscoelastic heat generation rate during welding.

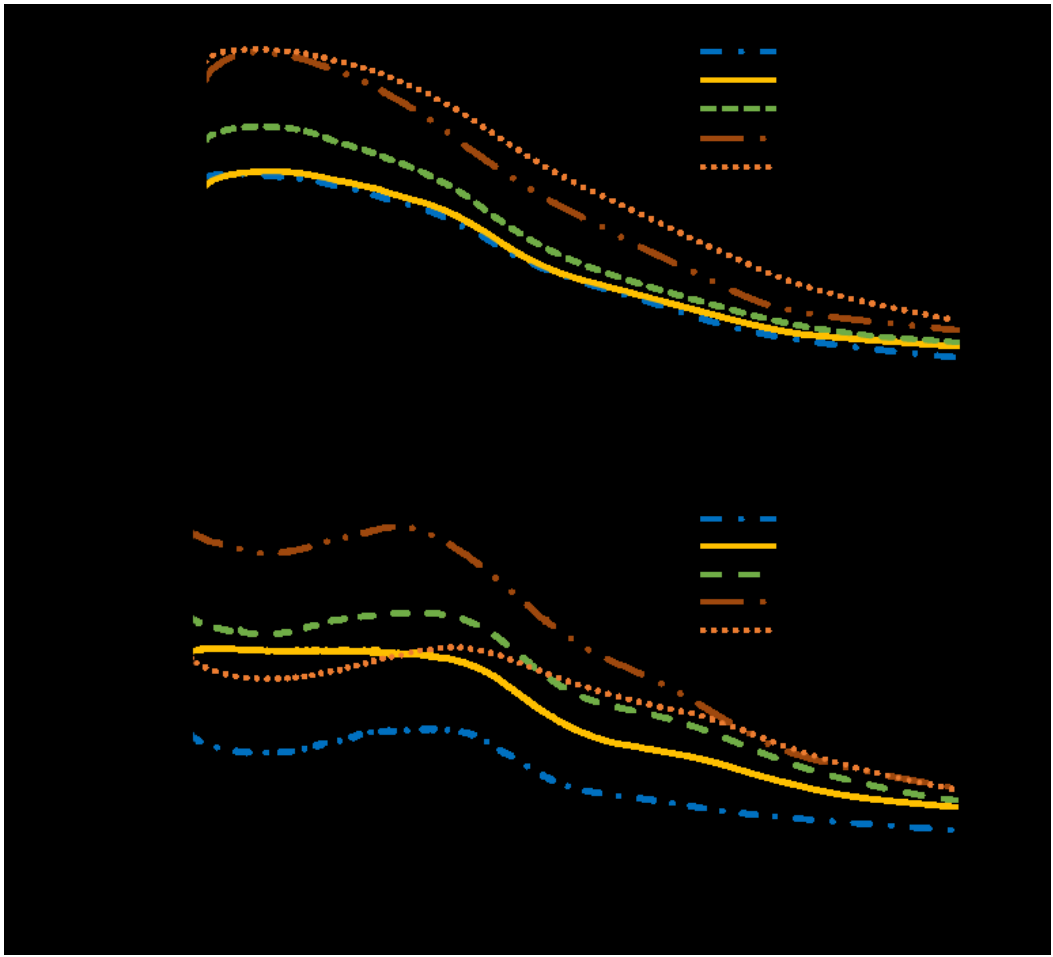

*Figure 6. (a) Storage modulus and (b) loss modulus of pure PP, 5 wt%, 10 wt%, 15 wt%, and 20 wt% MWCNT/PP films as a function of temperature.*

## 3.3 Effect of MWCNT films on Welding Process

Figure 7 plots the representative power and displacement curves of GF/PP joints welded with different nanocomposite films at 60% travel, namely pure PP, 5 wt% and 10 wt% MWCNT/PP, and embedded buckypaper films, where displacement represents the downward movement of the sonotrobe. These curves share a similar trend to samples welded with pure PP films, indicating negligible effect of multifunctional films on the welding process. Before 300 ms, films heat up and start to melt. During this phase, displacement curves are close to 0. As films continue to heat up, the displacement increases significantly due to the beginning of squeeze flow at the interface. Then, displacement continues to increase in a relatively slow trend associated with the melting and flow of the GF/PP adherends at the interface, or the flow of remaining films. The curves of the buckypaper film indicate a longer melting time, compared to all other films, because it is almost twice as thick (1.12 mm) as the others.

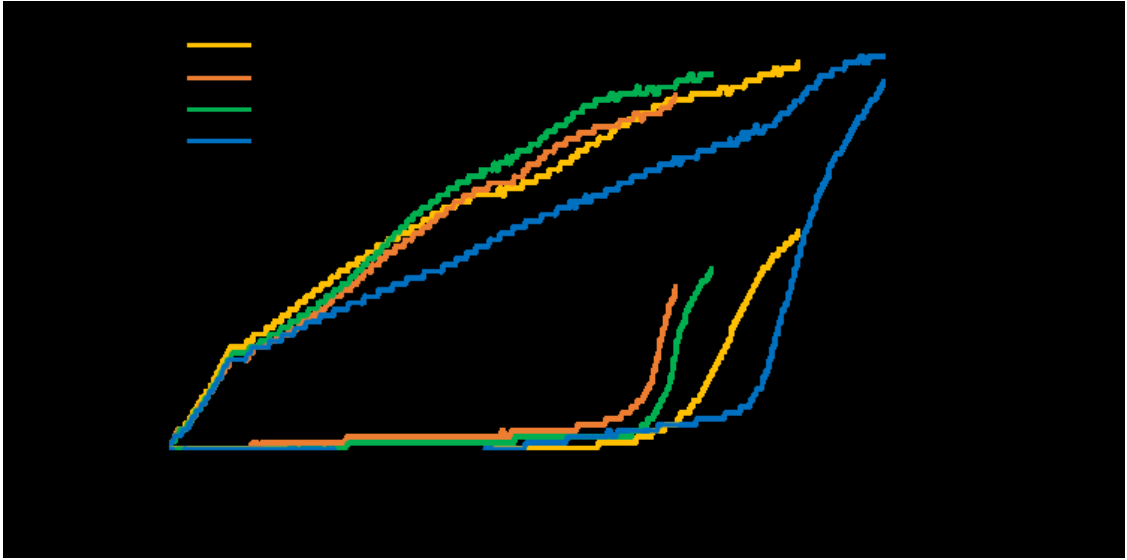

*Figure 7. Comparison of power and displacement curves for GF/PP joints welded with pure PP, 5 wt% and 10 wt% MWCNT/PP, and embedded buckypaper films at 60% travel.*

### 3.4 Electrical Resistance Response during Bending Tests

Figure 8 compares the resistance response of GF/PP joints welded with 5 wt%, 10 wt%, 15 wt%, and 20 wt% MWCNT/PP films, and embedded buckypaper as a function of time for a bending interval of 6 s. In Figure 8a, the resistance exhibits inconsistent changes under bending load, likely because there are few conductive paths at the interface due to the low MWCNT concentration (5 wt%). For 10 wt% MWCNT/PP films, shown in Figure 8b, the maximum and minimum resistance peaks also exhibit noise in the response data. However, films containing 15 wt% and 20 wt% MWCNT (Figure 8c-d) display the most consistent and repeatable behavior, capturing the gap increase between conductive MWCNT networks. However, for embedded buckypaper films, significantly higher resistance was measured, with no observable changes as load is applied (Figure 8e).

To explain the resistance behavior of the buckypaper film, Figure 9a-c compares the fracture surfaces of GF/PP joints welded with 5 wt% and 10 wt% MWCNT/PP films, as well as embedded buckypaper, respectively. Cohesive failure is observed on the surfaces of both adherends with melted MWCNT/PP films, presented in Figure 9a-b. There is no buckypaper on the fracture surfaces of all GF/PP joints (Figure 9c) due to its fragility leading to squeeze out during ultrasonic welding. This limits its application for ultrasonically welded joints. Therefore, it is necessary to develop a nanocomposite film, such as MWCNT/PP, that can sustain ultrasonic vibrations during the welding process for use as a SHM sensor.

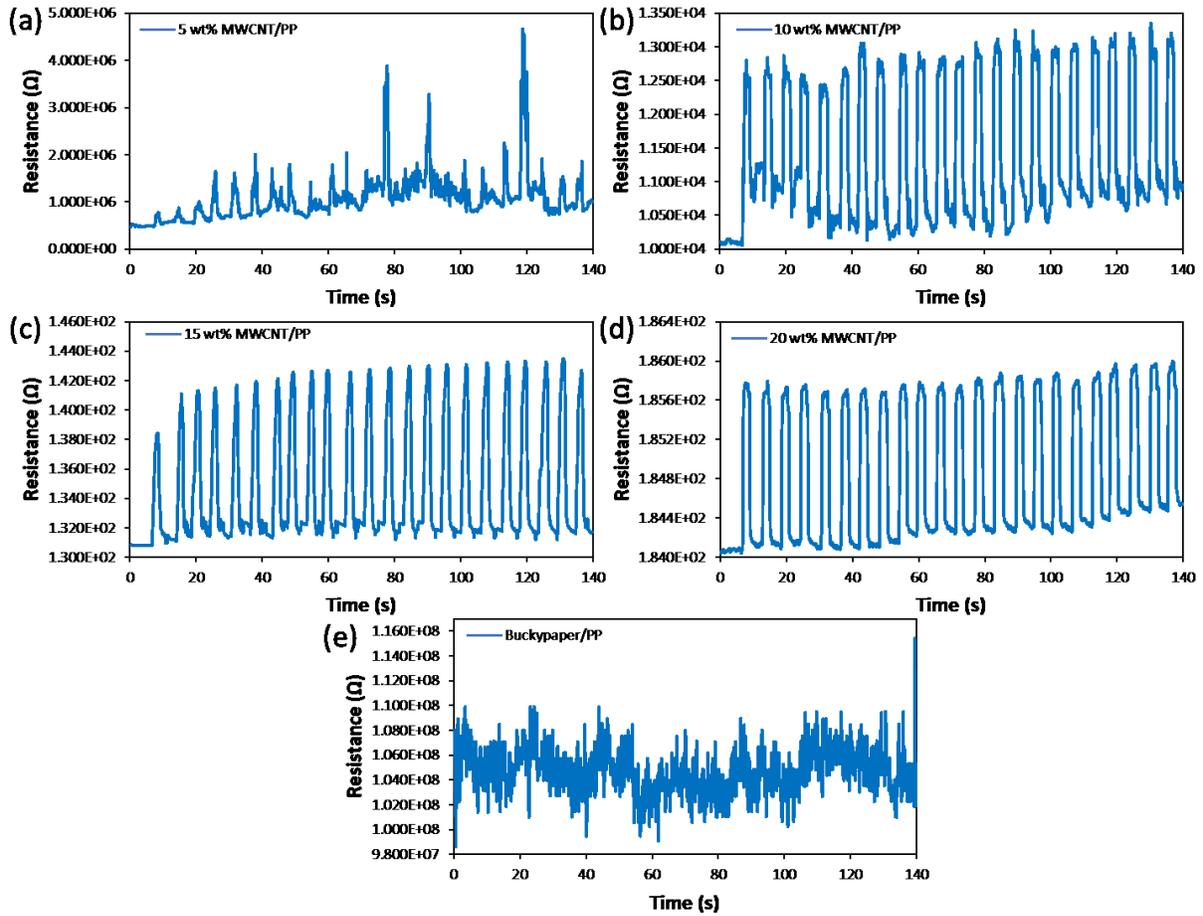

*Figure 8. Representative resistance curves of GF/PP samples welded with: (a) 5 wt%, (b) 10 wt%, (c) 15 wt%, and (d) 20 wt% MWCNT/PP films; and (e) embedded buckypaper at 60% travel as a function of time under periodic bending of 6 s.*

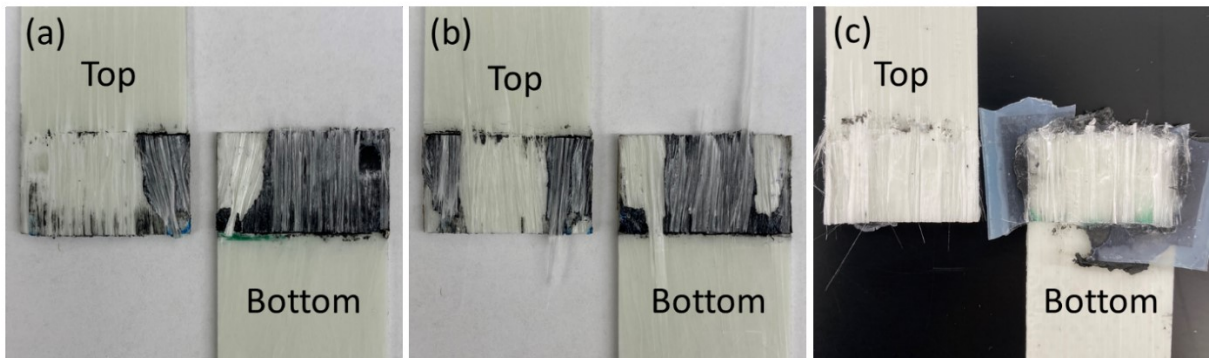

*Figure 9. Representative fracture surfaces of GF/PP joints welded with (a) 5 wt% MWCNT/PP film; (b) 10 wt% MWCNT/PP film; and (c) embedded buckypaper.*

As aforementioned, 15 wt% and 20 wt% MWCNT/PP films show higher stability and sensitivity than other films, which indicates that their application for SHM of TPC welded joints may be promising. In the next paragraph, 15 wt% MWCNT/PP films were chosen for further investigation.

A comparison of resistance responses for single lap GF/PP joints welded with 15 wt% MWCNT/PP films at 60% travel as a function of time under periodic bending at four different bending intervals (35 s, 20 s, 10 s, and 6 s) is presented in Figure 10a-d. The curves generally capture the intervals for all cases through repeatable resistance peak changes, which shows possibilities for SHM in real life applications. However, a shorter bending interval seems to lead to more stable readings for minimum and maximum peak values. This may be due to the fact that as the bending interval decreases, the load is applied again sooner before the resistance has time to change after load release.

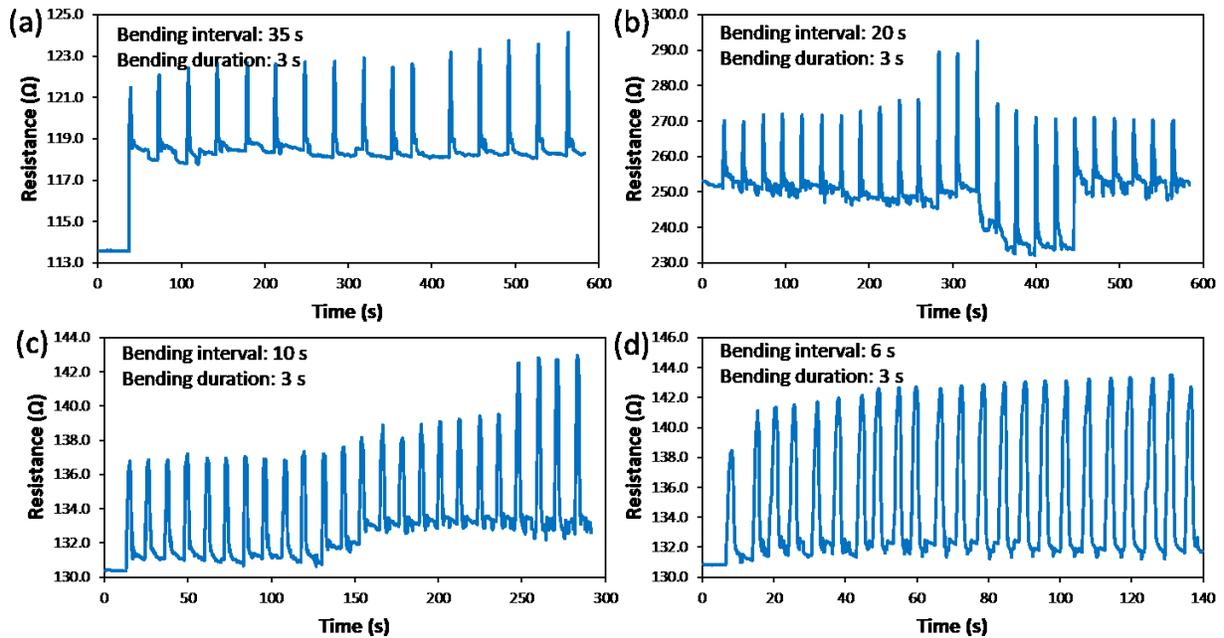

*Figure 10. Resistance response of GF/PP joints welded with 15 wt% MWCNT/PP film at 60% travel as a function of time under periodic bending for four bending intervals: (a) 35 s; (b) 20 s; (c) 10 s; and (d) 6 s.*

# 4. Conclusions

In this work, MWCNT/PP nanocomposites were synthesized to create films for ultrasonic welding and structural health monitoring. Hot pressing was performed to fabricate three types of multifunctional films. The effect of MWCNT concentration on the mechanical and electrical properties of nanocomposite films was examined by a series of experimental investigations. Moreover, the potential of multifunctional films for structural health monitoring was evaluated under bending loads. According to the above experiments, the following conclusions can be made:

- In general, under the same voltage, the average electrical conductivity increased with MWCNT weight fraction, except for 20 wt% MWCNT/PP films, possibly due to the poor nanotube dispersion. Moreover, electrical conductivity of MWCNT/PP films increased monotonically with increasing voltage, especially for films with high MWCNT concentration (15-20 wt%).

- Incorporation of MWCNTs can reinforce the matrix, improving storage and loss moduli. After glass transition temperature, the two moduli decrease as temperature increases.

- MWCNT/PP and embedded buckypaper films used for ultrasonic welding have no significant effect on welding process, which is consistent with our previous research [19].

- When periodic bending loads were applied to the welded joints, resistance curves captured the load intervals through repeatable peaks. Furthermore, multifunctional films containing 15 wt% and 20 wt% MWCNT displayed the most stable responses.

Overall, high weight fraction nanocomposite films (15 wt% or 20 wt% MWCNT) are recommended for structural health monitoring of thermoplastic composite joints due to their excellent electrical and mechanical properties, as well as high sensitivity and stability. Future work will focus on i) investigating the dispersion state of nanofiller within polymer matrix, and ii) developing data analysis methods to relate resistance changes to specific loading cases and failure modes. Finally, while commercially available MWCNT masterbatches are relatively low cost, as a next step, the incorporation of carbon black (CB) with lower MWCNT weight fractions may lead to more cost efficient, USW-compatible films.